# A linear complexity analysis of quadratic residues and primitive roots spacings

Mihai Caragiu†, Shannon Tefft, Aaron Kemats and Travis Maenle

Department of Mathematics and Statistics, Ohio Northern University

**ABSTRACT**

We investigate the linear complexities of the periodic 0 – 1 infinite sequences in which the periods are the sequence of the parities of the spacings between quadratic residues modulo a prime p, and the sequence of the parities of the spacings between primitive roots modulo p, respectively. In either case, the Berlekamp-Massey algorithm running on MAPLE computer algebra software shows very good to perfect linear complexities.

**2010 MSC:** 11B50, 11Y16, 11T71

**Keywords and phrases:** *linear complexity, quadratic residues, primitive roots, spacings, computational number theory*

______________________________

†Corresponding author

## 1. Introduction

The linear complexity of quadratic residues has been exactly calculated. One could customarily associate to a prime $p$ the infinite periodic "Legendre sequence" $\{x_n\}_{n\geq 0}$ of period $p$ defined as follows:

$$(1) \quad x_n = \begin{cases} \left[1 + \left(\dfrac{n}{p}\right)\right]\Big/2, & \text{if } n \not\equiv 0 \bmod p \\ 0, & \text{else} \end{cases}$$

It is proved [3] that the linear complexity of $\{x_n\}_{n\geq 0}$ (the length of the shortest linear feedback shift register generating $\{x_n\}_{n\geq 0}$) can be explicitly calculated as follows:

$$L(\{x_n\}_{n\geq 0}) = \begin{cases} (p+1)/2, & \text{if } p = 8t - 1 \\ (p-1)/2, & \text{if } p = 8t + 1 \\ p, & \text{if } p = 8t + 3 \\ p - 1, & \text{if } p = 8t + 5 \end{cases}$$

Consequently, for $p \equiv \pm 3 \bmod 8$, the linear complexity of the Legendre sequence (1) is perfect or almost perfect. This is consistent with the fact that periodic sequences with periods generated by independent and uniformly distributed binary random variables are good on average [7].

In the present sequence we investigate two linear complexity measures associated to important number theoretic sequences, namely the sequence of spacings between quadratic residues, and the sequence of spacings between consecutive primitive roots.

Let the quadratic residues modulo $p$ be $1 = x_1 < x_2 < ... < x_{(p-1)/2}$. The quadratic residue spacings form the following list with $(p-3)/2$ elements:

$$x_2 - x_1, x_3 - x_2, ..., x_{(p-1)/2} - x_{(p-3)/2}$$

It is known that the quadratic residue spacings have an exponential distribution, with the number of pairs of consecutive quadratic residues at distance 2 being roughly half the number of pairs of consecutive quadratic residues with distance 1, etc. This exponential distribution result goes back to Davenport [2]. The binary sequences we are interested in connection to quadratic residues modulo prime numbers consist of the parities of quadratic residue spacings:

(2) $Q_p := \left[ x_2 - x_1 \mod 2, x_3 - x_2 \mod 2, ..., x_{(p-1)/2} - x_{(p-3)/2} \mod 2 \right]$

We will use the Berlekamp-Massey algorithm to compute the linear complexities of infinite sequences $Q_p^*$ with the binary lists (2) as periods (that is, $Q_p^*$ is an infinite concatenation of blocks identical to $Q_p^*$). Calculating $L(Q_p^*)$ can be done by applying the Berlekamp – Massey algorithm to the duplicated list $[Q_p, Q_p]$. This is because the Berlekamp-Massey algorithm invariably provides the same, uniquely determined linear complexity of a linear shift register sequence if applied to an initial segment of bits of length at least twice the period ([1] section 5.4., [5], [6]). Hence, procedurally, as a linear complexity measure for $Q_p^*$ we will use the normalized quantity representing a value $c_p \in [0,1]$:

(3) $c_p = \dfrac{L([Q_p, Q_p])}{(p-3)/2}$.

We will take a similar approach with primitive roots (generators of the multiplicative group) modulo a prime. Let the primitive roots modulo $p$ be $g_1 < g_2 < ... < g_{\varphi(p-1)}$ where $\varphi$ represents the Euler's phi function. As in the case of quadratic residues, the arithmetical object of interest consists of the $\varphi(p-1)-1$ spacings between consecutive primitive roots:

$g_2 - g_1, g_3 - g_2, ..., g_{\varphi(p-1)} - g_{\varphi(p-1)-1}$

The primitive roots spacings also satisfy an exponential distribution, as shown by Zaharescu and Cobeli [8]. In this case, the binary sequences of interest consist of the parities of primitive roots spacings

$$(4)\ R_p := \left[ g_2 - g_1 \mod 2, g_3 - g_2 \mod 2, ..., g_{\varphi(p-1)} - g_{\varphi(p-1)-1} \mod 2 \right].$$

As before, $R_p$ is the period of an infinite sequence $R_p^*$, for which a linear complexity measure $d_p \in [0,1]$ will be obtained by applying the Berlekamp-Massey algorithm to the duplicate $\left[ R_p, R_p \right]$ followed by a suitable normalization:

$$(5)\ d_p = \frac{L\left(\left[R_p, R_p\right]\right)}{\varphi(p-1)-1}.$$

The case in which $c_p, d_p$ (associated to their corresponding bit strings with periods $(p-3)/2$ and $\varphi(p-1)-1$, respectively) are close to 1 is precisely the "good linear complexity" case, with the concrete interpretation: to convey or communicate such a periodic bit string, we cannot do much better than simply listing the full period.

## 2. Computational results

In our computations, we used a MAPLE version of the Berlekamp-Massey algorithm [4] in the form of the procedure *BM(s, N, p, x)* that produces the feedback polynomial in the variable *x*, for a sequence *s* of minimum length *2N*, with terms in a field of characteristic *p*. The degree of *BM(s, N, p, x)* will be used as a convenient measure in our linear complexity calculations (in which typically we choose $p = 2$).

To get the linear complexities of the parities of quadratic residue spacings corresponding to the 1000 primes in a row beginning with 5 and ending at 7933 we used a simple MAPLE code. The idea is simple: for any prime p in the desired range,

- Calculate the squares of the elements from 1 to $(p – 1)/2$ modulo p
- Sort the list of squares
- Calculate the differences between consecutive terms of the sorted list (these would be the list spacings, with $(p – 3)/2$ elements)
- Reduce the list of spacings modulo 2 then duplicate it
- Apply the Berlekamp-Massey algorithm to the duplicated binary list (the output is the degree of the feedback polynomial)
- Normalize the linear complexity output by dividing it by $(p – 3)/2$, thus getting the complexity measure $c_p \in [0,1]$.

At the end, the sequence of complexities $\{c_p\}$ is analyzed through a histogram or related frequencies counting. The MAPLE code introduced above will look as follows:

*N:=1000: for r from 1 to N do p:=ithprime(r+2): for k from 1 to (p-1)/2 do y(k):=k^2 mod p end do: Y:=[seq(y(k),k=1..(p-1)/2)]: X:=sort(Y): Z:=[seq(X[s+1]-X[s],s=1..(p-3)/2)]: QR:=[op(Z mod 2), op(Z mod 2)]: C(r):=evalf(degree (BM(QR,(p-3)/2,2,x))/((p-3)/2)): end do: LC:=[seq(C(r),r=1..N)]; Histogram(LC, frequencyscale=absolute);*

The result of the MAPLE analysis shows very good to perfect linear complexities in the 1000 elements list $(c_p)_{5 \leq p \leq 7933}$ corresponding to the 1000 primes starting with 5 (the starting point is chosen so that at least two quadratic residues exist). Out of a total of 1000 items, we find

- 671 perfect linear complexities of 1
- 991 linear complexities of at least 0.95
- 967 linear complexities of at least 0.99

For a better visibility, Figure 1 below displays only the subset of 967 elements of $(c_p)_{5 \leq p \leq 7933}$ that are 0.99 or larger.

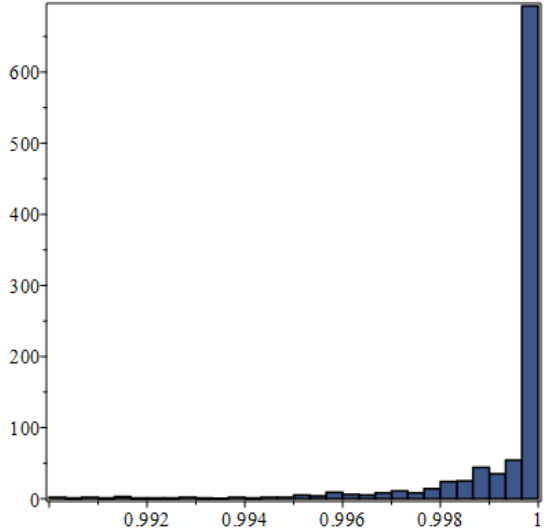

**Figure 1. Visual display of the top 967 out of the 1000 linear complexities $(c_p)_{5 \leq p \leq 7933}$ defined by the quadratic residue spacing parities.**

To get the linear complexities of the parities of primitive root spacings corresponding to the 1000 primes in a row beginning with 11 and ending at 7949 we used a simple MAPLE code. The idea is similar: for any prime $p$ in the desired range,

- Produce a complete list of primitive roots modulo p starting from the least primitive root primroot(p) – the first term of the list - and recursively appending the least primitive root greater than the previous term in the list. The list is already sorted and has $\varphi(p-1)$ elements

- Calculate the differences between consecutive terms (these would be the list of primitive roots spacings, with $\varphi(p-1)-1$ elements)

- Reduce the list of spacings modulo 2 then duplicate it

- Apply the Berlekamp-Massey algorithm to the duplicated binary list (the output is the degree of the feedback polynomial)

- Normalize the linear complexity output by dividing it by $\varphi(p-1)-1$, thus getting the corresponding complexity measure $d_p \in [0,1]$.

The MAPLE code that we used is shown below.

*N:=1000: for r from 1 to N do p:=ithprime(r+5); g:=primroot(p): x(1):=g: for k from 2 to phi(p-1) while x(k) <>1 do x(k):=primroot(x(k-1),p) end do: L:=[seq(x(k),k=1..phi(p-1))]: M:=[seq(x(k+1)-x(k),k=1..phi(p-1)-1)]: PRR:=[op(M mod 2),op(M mod 2)]: E(r):=evalf(degree(BM(PRR, phi(p-1)-1,2,x),x)/(phi(p-1)-1)): end do: LR:=[seq(E(r),r=1..N)]; Histogram(LR, frequencyscale=absolute);*

The result of the MAPLE analysis of the primitive roots spacings shows, again, very good to perfect linear complexities in the 1000 elements list $(d_p)_{11 \leq p \leq 7949}$ corresponding to the 1000 primes starting with 11 (the starting point is chosen so that at least two primitive roots exist). Out of a total of 1000 items, we find

- 610 perfect linear complexities of 1
- 980 linear complexities of at least 0.968
- 953 linear complexities of at least 0.992

For a better visibility, Figure 2 below displays only the subset of 967 elements of $(d_p)_{11 \leq p \leq 7949}$ that are 0.992 or larger.

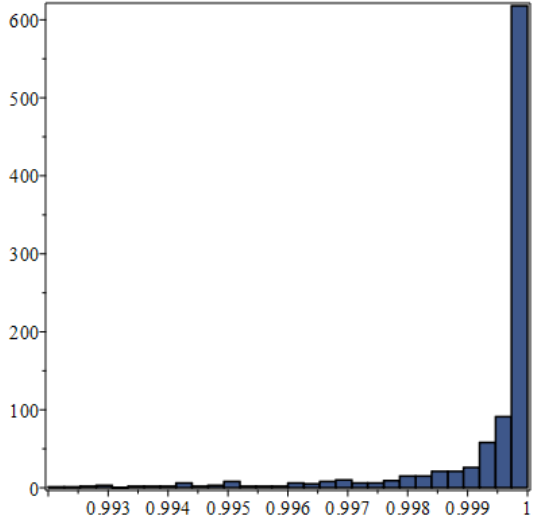

**Figure 2. Visual display of the top 953 out of the 1000 linear complexities $(d_p)_{11 \leq p \leq 7949}$ defined by the primitive roots spacing parities.**

## 3. Monte-Carlo experiments with quadratic residues and primitive roots spacings

In what follows we will randomize the search for complexities of the parities of quadratic residues and primitive root spacings, and the result will be a Monte Carlo type analysis in which, for each of the $N$ randomly selected large primes $p$, we look at the sequences of quadratic residues and primitive roots within a randomly selected discrete interval of size $K$ in the set of integers modulo $p$. Within any such interval $I$ (which may be seen as a "window" of size $K$ into the larger total set of quadratic residues or primitive roots modulo $p$, respectively) we perform essentially the same type of analysis as the one effected in Section 2 above.

Here are some specifications and outcomes of our Monte Carlo – type experiments.

**Quadratic residues case**

- $N$ random selections of prime numbers $p_r$ with $10^{30} < p_r < 10^{40}$, $1 \leq r \leq N$ are made.
- Within any such finite field $GF(p_r)$ we randomly select an interval $I_r \subset GF(p_r)$ with $|I_r| = K$.
- Note that in the sample calculation presented here, we use $N = K = 1000$.

- The differences between consecutive quadratic residues within the interval $I_r$ are then calculated, and the differences are then reduced modulo 2 to produce a binary string $W_r$.

- The Berlekamp-Massey algorithm calculates the linear complexity of the infinite periodic bit stream having $W_r$ as period, which is normalized by dividing it by $|W_r|$, thus producing an element $\text{QRCOMPL}(r) \in [0,1]$.

- A histogram of list consisting of $\text{QRCOMPL}(r), 1 \leq r \leq N$ is produced and the corresponding tally counts are analyzed. A typical such histogram is presented in Figure 3 below: out of 1000 linear complexities calculated, 947 are at least 0.99158 while 486 are of a perfect linear 1.0 complexity.

The MAPLE code corresponding to the quadratic residue spacings analysis is presented below:

```
K:=1000: N:=1000: for r from 1 to N do p:=nextprime(rand(10^30..10^40)()):
A:=Generate(integer(range=1..p-K)): for k from 1 to K do x(k):=(1+legendre(A+k-1,p))/2 end do:
L:=[seq(x(k),k=1..K)]: M:=select(i->L[i]=1, [$ 1..nops(L)]); W:=[seq(M[t+1]-M[t],t=1..numelems(M)-1)]: QR:=[op(W mod 2),op(W mod 2)]: QRCOMPL(r):=evalf(degree(BM(QR,numelems(M)-1,2,x),x)/(numelems(M)-1)): end do: QRLIST:=[seq(QRCOMPL(r),r=1..N)]:
Histogram(QRLIST,frequencyscale=absolute);
```

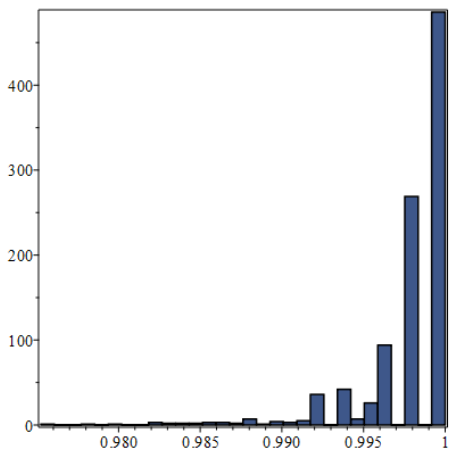

**Figure 3. Randomized exploration – linear complexities of quadratic residues spacing parities, for intervals of size 100 mod $p$, with $10^{30} < p < 10^{40}$**

**Primitive Roots case**

This follows closely the ideas outlined above in the quadratic residues analysis, but working instead with the primitive roots in the randomly selected interval $I_r$ of the randomly selected finite field $GF(p_r)$, thus producing a list of normalized linear complexities $\text{PRCOMPL}(r), 1 \leq r \leq N$. A histogram of list consisting of $\text{PRCOMPL}(r), 1 \leq r \leq N$ is produced and the corresponding tally counts are analyzed. A typical such histogram is presented in Figure 4 below: out of 1000 linear complexities calculated, 901 are at least 0.99057, while 465 are of a perfect 1.0 complexity.

The MAPLE code corresponding to the primitive roots spacings analysis is presented below:

*K:=1000: N:=1000: for r from 1 to N do p:=nextprime(rand(10^30..10^40)()):*
*A:=Generate(integer(range=1..p-K)): for k from 1 to K do x(k):=order(A+k-1,p) end do:*
*L:=[seq(x(k),k=1..K)]: M:=select(i->L[i]=p-1, [$ 1..nops(L)]); W:=[seq(M[t+1]-M[t],t=1..numelems(M)-1)]: PR:=[op(W mod 2),op(W mod 2)]:*
*PRCOMPL(r):=evalf(degree(BM(PR,numelems(M)-1,2,x),x)/(numelems(M)-1)): end do:*
*PRLIST:=[seq(PRCOMPL(r),r=1..N)]: Histogram(PRLIST,frequencyscale=absolute);*

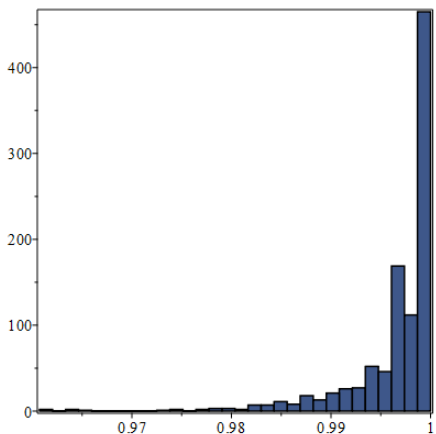

**Figure 4. Randomized exploration – linear complexities of primitive roots spacing parities, for intervals of size 100 mod $p$, with $10^{30} < p < 10^{40}$**

## 4. Conclusions

The objective of the present article was to draw the attention to the consistently good linear complexities of binary sequences with periods consisting of parities of quadratic residues and primitive roots spacings modulo prime numbers, with potential use in cryptography – related applications. In the randomized search, for an increased size of the interval, a substantial increase in the running time has been noticed, but the results were largely similar. For example, in an analysis of quadratic residue spacings in the same vein, with N random selections of discrete intervals $I_r$ with $|I_r| = 10000$ over a randomized selection of primes $10^{30} < p < 10^{40}$ in the output in the form of 1000 linear complexities, 994 of them were of at least 0.99854 and 486 were of a perfect 1.0. The more tedious computations have been performed with MAPLE 2016 on a 12 GB RAM Intel(R) Core(TM) i7-6500U CPU @ 2.50GHz platform. The study was done in the context of the 1-credit class "Experimental Number Theory" at Ohio Northern University in the fall semester of 2018.

CORRESPONDING AUTHOR'S ADDRESS:

Mihai Caragiu
Department of Mathematics and Statistics
Ohio Northern University
525 S. Main Street, Ada, OH 45810